\documentclass[twocolumn,showpacs,preprintnumbers,amsmath,amssymb]{revtex4}
\usepackage{graphics}
\usepackage{graphicx}

\begin{document}

\title{Entanglement manipulation by a local magnetic pulse}

\author{Shi-Jian Gu}
\author{Hai-Qing Lin}
\affiliation{Department of Physics, The Chinese University of Hong Kong, Hong
Kong, China}

\date{\today}

\begin{abstract}
A scheme for controlling the entanglement of a two-qubit system by a local
magnetic pulse is proposed. We show that the entanglement of the two-qubit
system can be increased by sacrificing the coherence in ancillary degree of
freedom, which is induced by a local manipulation.
\end{abstract}
%03.65.Ud
%Entanglement and quantum nonlocality (e.g. EPR paradox, Bell's
%inequalities, GHZ states, etc.) (for entanglement production in quantum
%information, see 03.67.Mn; for entanglement in Bose-Einstein condensates, see
%03.75.Gg)

%03.67.Mn Entanglement production, characterization, and manipulation (see also
%03.65.Ud Entanglement and quantum nonlocality; for entanglement in
% Bose-Einstein condensates, see 03.75.Gg)

%76.90.+d Other topics in magnetic resonances and relaxations (restricted to new
%topics in section 76)

\maketitle

Quantum entanglement plays a central role in quantum information and quantum
computation \cite{Nielsen1,AGalindo02}. It is in the heart of most quantum
phenomena, such as quantum teleportation, dense coding, and quantum
cryptography \cite{Bennett}. Thus, the issue of creating two-particle
entanglement in various composite systems is of great importance in technology
development. For example, in nuclear magnetic resonance experiments
(NMR)\cite{LMKVandersypen04,ILChuang98,SLloyd98,RJNelson00}, significant
efforts were put in creating Bell states and Greenberger- Horne-Zeilinger (GHZ)
state. Schemes for producing entangled state using adiabatic population
transfer based on a two-spin system have also been
proposed\cite{RGUnanyan01,BZhou04}. Quite recently, it has been shown that the
existence of relative phase between two Rabi frequencies can be used to control
entanglement of the two-qubit system at will\cite{VSMalinovsky04}.

As it is well known, the entanglement is intrinsically related to the
superposition principle of quantum mechanics. Thus the problem of creating or
controlling the entanglement is simply the problem of coherent control of
population transfer between different levels of a composite system. For
example, the maximally entangled state $(|00\rangle+|11\rangle)/\sqrt{2}$ can
be created from a fully polarized state $|00\rangle$ by transferring half of
its population to state $|11\rangle$ through a careful controlled adiabatic
process: $|00\rangle\rightarrow |01\rangle \rightarrow |11\rangle$. A
successful realization of population transfer depends on the adiabatic
condition and can be achieved through a careful control of pulse shape with an
appropriate choice of the Rabi oscillation frequency\cite{PMarte91,MWeitz94}.

On the other hand, in order to be useable for quantum information process, one
particle of the prepared entangled state has to be sent through a classical
channel. Then the entanglement between the two particles, in general, will be
weakened because of its interference with the environment. However, in some
purposes such as teleportation, the entanglement must be supplied in the form
of maximally entangled pairs. Therefore the entanglement
distillation\cite{Bennett1} is of importance in quantum information. Through a
method called Schmidt projection, it has been shown that the entanglement of
$n$ partly entangled states can be concentrated into a small number of
maximally entangled pairs\cite{CHBennett96}.

In this paper, we propose a method to control the entanglement of two separated
systems by modulating a local oscillating field which is used to create
coherent superposition in one of the two systems. We show, as long as the
entanglement is non-zero, the entanglement can be modulated to reach a maximum
with a certain probability via an appropriate choice of tuning period and radio
magnitude. Therefore, we can choose the pulse strength or the tuning period or
both as the control knob to create the state with desired entanglement,
especially the maximally entangled state. As we will show below, the external
oscillation in our method only interacts with one of the particle and can be
controlled locally. This method is different from the Schmidt
projection\cite{CHBennett96}. Therefore, the suggested method opens a new
avenue for entanglement distillation and facilitates experimental
implementation on quantum information transmission.

{\it Entanglement evolution and projection:} We consider two separated
non-interacting particles called A and B, shared between Alice and Bob
respectively. In addition to the spin degree of freedom which are partly
entangled between A and B, such as the state
$\cos(\theta)|\uparrow\uparrow\rangle+\sin(\theta)|\downarrow
\downarrow\rangle$, the system has other ancillary degree of freedom denoted as
$|0\rangle$ and $|1\rangle$. This ancillary degree of freedom might be the
orbital state in spin-orbit interacting systems, energy level of quantum dot,
or level of ions in magnetic trap. For convenience, we call it band hereafter.
Therefore, there are four possible local states, i.e., $|0\uparrow\rangle$,
$|1\uparrow\rangle$, $|0\downarrow\rangle$, and $|1\downarrow\rangle$, either
for particle A or particle B. In a magnetic field, the Hamiltonian of particle
A and particle B can be written as
\begin{eqnarray}
H_0^{A(B)}=-[\omega_s\sigma^z + \omega_b \tau^z - J\sigma^z\tau^z],
\end{eqnarray}
which is typical in the NMR systems\cite{LMKVandersypen04}. Here $\omega_s,
\omega_b$ are the Larmor frequency for spin ($\sigma^z$) and band ($\tau^z$)
respectively, while the scalar coupling $J$ can be interpreted as the coupling
from band to an additional static field along $z$ direction produced by spin
(or vice versa). Clearly, the Hamiltonian is diagonal in the standard basis,
and the corresponding eigenvalues are
\begin{eqnarray}
&&\varepsilon_1=-\omega_s - \omega_b + J, \;\;\varepsilon_2=-\omega_s +
\omega_b
- J, \nonumber \\
&&\varepsilon_3=\omega_s - \omega_b - J, \,\;\;\;\;\varepsilon_4=\omega_s +
\omega_b + J.
\end{eqnarray}

\begin{figure}
\includegraphics[width=8cm]{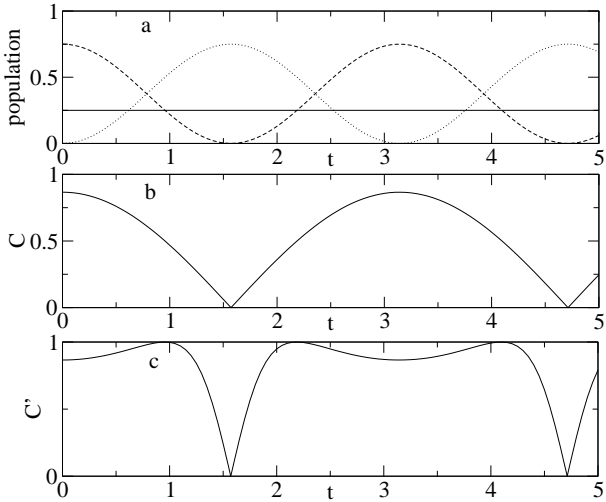}
\caption{(a) Population dynamics of the state $|00\rangle|\downarrow
\downarrow\rangle$ (solid line), $|00\rangle |\uparrow\uparrow\rangle$ (dashed
line) and $|01\rangle |\uparrow\uparrow\rangle$ (dotted line) for the case of
$\theta=\pi/6$. (b) The evolution of the concurrence between the two spins. (c)
The evolution of the concurrence if the projection measurement is successful.
Here, the time is in unit of $1/g$.} \label{fig:dynaent}
\end{figure}

Since we are interested in the entanglement between the two spins, it is
assumed that the two spins in the initial state are not maximally entangled,
while two bands are fully polarized. Mathematically, we assume that the initial
state takes the form
\begin{eqnarray}
\Psi(t=0)=\cos\theta |0\uparrow\rangle_A|0\uparrow\rangle_B+\sin\theta
|0\downarrow\rangle_A |0\downarrow\rangle_B.\label{eq:initalstate}
\end{eqnarray}
Here the partly entangled state is not restricted to the form $\cos\theta
|\uparrow\uparrow\rangle+ \sin \theta|\downarrow\downarrow\rangle$, it can be
in other forms, such as $\cos\theta |\downarrow\uparrow\rangle- \sin
\theta|\downarrow\uparrow\rangle$. The entanglement of two spins in this state,
measured by the concurrence\cite{WKWootters98}, is $C(t=0)=|\sin 2\theta|$.
Thus the first problem is to find a possible magnetic pulse that leads to a
unitary time evolution, and then to modulate the population of
$|\uparrow\rangle_A|\uparrow\rangle_B$ and
$|\downarrow\rangle_A|\downarrow\rangle_B$. For this purpose, we add an ideal
rectangular transversal magnetic pulse with frequency $\omega$ on particle B.
This additional Hamiltonian can be simplified as
\begin{eqnarray}
H^B_R=g [e^{i\omega t + i\phi}\tau^\dagger + h.c.]
\end{eqnarray}
where $\tau^+|0\rangle_B = |1\rangle_B,\, \tau^-|1\rangle_B = |0\rangle_B$,
$\phi$ is the relative phase and $g$ is the magnitude of the magnetic pulse.
Thus the whole Hamiltonian of the two-particle system is $$H=H^A_0 +H^B_0 +
H_R^B,$$ and the time-dependent Schr\"{o}dinger equation reads
\begin{eqnarray}
H \Psi(t)=i\frac{i\partial}{\partial t} \Psi(t).
\end{eqnarray}
In order to eliminate the time dependence of the Hamiltonian of particle B:
$H^B=H^B_0 + H_R^B$, we apply a unitary transformation
\begin{eqnarray}
U &=& e^{-i\omega t} |0\uparrow\rangle_{BB}\langle 0\uparrow| + e^{i\omega t}
|1\uparrow\rangle_{BB}\langle 1\uparrow| \nonumber \\ &+& e^{-i\omega t}
|0\downarrow\rangle_{BB}\langle 0\downarrow| + e^{i\omega t}
|1\downarrow\rangle_{BB}\langle 1\downarrow|.
\end{eqnarray}
Then the effective ``rotating frame" Hamiltonian matrix of particle B becomes
\begin{eqnarray}
&& H^B_U=UH^B U -iU\frac{\partial}{\partial t}U^\dagger\\ \nonumber  &&=\left(%
\begin{array}{cccc}
  \varepsilon_1^B + \omega/2 & g e^{i\phi} & 0 & 0 \\
  g e^{-i\phi} & \varepsilon_2^B - \omega/2 & 0 & 0 \\
  0 & 0 & \varepsilon_3^B + \omega/2 & g e^{i\phi} \\
  0 & 0 & g e^{-i\phi} & \varepsilon_4^B - \omega/2
\end{array}%
\right)
\end{eqnarray}
in the standard basis $|0\uparrow\rangle_B, |1\uparrow\rangle_B,
|0\downarrow\rangle_B, |1\downarrow\rangle_B$ of particle B. The eigenvalues of
$H^B_U$ are
\begin{eqnarray}
E_{1(2)}=\frac{1}{2}\left[(\varepsilon_1^B+\varepsilon_2^B)\pm\sqrt{(\varepsilon_1^B
-\varepsilon_2^B+\omega)^2 + 4g^2}\right]\\
E_{3(4)}=\frac{1}{2}\left[(\varepsilon_3^B+\varepsilon_4^B)\pm\sqrt{(\varepsilon_3^B
-\varepsilon_4^B+\omega)^2 + 4g^2}\right]
\end{eqnarray}
and the corresponding eigenvectors are
\begin{eqnarray}
\psi_{1(2)}=|0\uparrow\rangle_B+
e^{-i\phi}\left(\pm\sqrt{1+\eta_{12}^2}-\eta_{12}\right) |1\uparrow\rangle_B, \\
\psi_{3(4)}=|0\downarrow\rangle_B+
e^{-i\phi}\left(\pm\sqrt{1+\eta_{34}^2}-\eta_{34}\right) |1\downarrow\rangle_B,
\end{eqnarray}
where $\eta_{ij}=(\varepsilon^B_i - \varepsilon^B_j +\omega)/2g$ is a
dimensionless parameter. We choose
$\omega=\varepsilon_2-\varepsilon_1=2\omega_b-2J$ and take the weak field
limit, i.e., $\eta_{34}\gg 1$. Then the dynamics of the whole system is
dominated by the resonance between $|0\uparrow\rangle_B$ and
$|1\uparrow\rangle_B$ in system B. Therefore, to the order of $O(1/\eta_{34})$,
the wave function of the two particles can be expressed as
\begin{eqnarray} \nonumber
\Psi(t)&=& c_0(t)|0\downarrow\rangle_A|0\downarrow\rangle_B + c_1(t)
|0\uparrow\rangle_A|0\uparrow\rangle_B \nonumber \\ &&+c_2(t)
|0\uparrow\rangle_A|1\uparrow\rangle_B,
\end{eqnarray}
with the amplitude $c_i(t), i=1,2,3$ as
\begin{eqnarray}
&&c_0(t)=e^{-i(\varepsilon_3^A+\varepsilon_3^B)t}\sin\theta,\nonumber \\
&&c_1(t)=e^{-i(\varepsilon_1^A+\varepsilon_1^B) t}\cos(gt)\cos\theta, \\
&&c_2(t)=e^{-i(\varepsilon_1^A+\varepsilon_2^B) t}\sin(gt)\cos\theta
e^{-i(\phi+\pi/2)}. \nonumber
\end{eqnarray}
Therefore the magnetic pulse introduces a resonance for particle B via an
appropriate choice on the frequency, and the resonance depresses the amplitude
of $|0\uparrow\rangle_A|0\uparrow\rangle_B$. Obviously, the population transfer
here is selective and the resonance can be regarded as a filter\cite{DYang04}
in the process of population transfer. We show that the population dynamics of
the corresponding frequency in Fig. \ref{fig:dynaent}(a) for the case of
$\theta=\pi/6$. Obviously, the entanglement of the two spins is also suppressed
by the local magnetic pulse, and it evolves in the form of
\[ C(t) = |\sin(2\theta)\cos(gt)| = C(t=0) |\cos(gt)| \]
as time elapses. This is consistent with the fact that the local operation and
classical communication can not increase the entanglement between two parties.

In order to increase the entanglement between particle A and B, we now perform
a following projection measurement on particle B,
\begin{eqnarray}
P_0=|0\rangle_{BB}\langle 0|,\; P_1=|1\rangle_{BB}\langle 1|.
\label{eq:projectionoperator}
\end{eqnarray}
After the measurement, the band of particle B will be projected onto state
$|0\rangle_B$ and $|1\rangle_B$ with probability $\sin^2\theta +
\cos^2\theta\cos^2(gt)$ and $\cos^2\theta \sin^2(gt)$, respectively. Clearly,
the latter case makes no sense because the entanglement between two particles
is completely destroyed. We are only interested in the former case in which the
output state becomes
\begin{eqnarray}
\Psi'(t)=\frac{\left[c_0(t)|0\downarrow\rangle_A|0\downarrow\rangle_B + c_1(t)
|0\uparrow\rangle_A|0\uparrow\rangle_B\right]}{\sqrt{c_0(t)^2+c_1(t)^2}}.
\end{eqnarray}
This state may possess higher entanglement than the original state, as
reflected by its concurrence measure
\begin{eqnarray}
C'(t)=\frac{\sin(2\theta)\cos(gt)}{\sin^2\theta + \cos^2\theta\cos^2(gt)}
\end{eqnarray}
which reaches maximum $C'=1$ at the condition $\cos(gt)=\tan\theta$, as is
shown in Fig. \ref{fig:dynaent}(c). This properties is completely different
from the behavior of $C(t)$, which is always suppressed during the evolution.
The reason we have such behavior of $C'(t)$ is due to the superposition
principle of quantum mechanics. Therefore, Bob could tell Alice in a classical
way whether his projection measurement is successful or not and only when the
output of his measurement is $|0\rangle_B$, the final state is entangled,
otherwise it is not entangled.

In short, we summarize the whole procedure as follows
\begin{itemize}
\item First, if Alice and Bob initially share a partly entangled state, such as
$\cos\theta |\uparrow\rangle_A|\uparrow\rangle_B+\sin\theta
|\downarrow\rangle_A |\downarrow\rangle_B$, then Bob will switch on a magnetic
pulse to transfer the population of $|\uparrow\rangle_A|\uparrow\rangle_B$ to a
target state which is distinguished by introducing an ancillary degree of
freedom.

\item Second, Bob does a projection measurement in the space of ancillary
degree of freedom.

\item Finally, Bob tells Alice his projection measurement result through a
classical channel.
\end{itemize}

This completes the procedure for a single pair. Clearly, the third step is
quite necessary since the final result depends on the probability of the
projection measurement (\ref{eq:projectionoperator}).

\begin{figure}
\includegraphics[width=8cm]{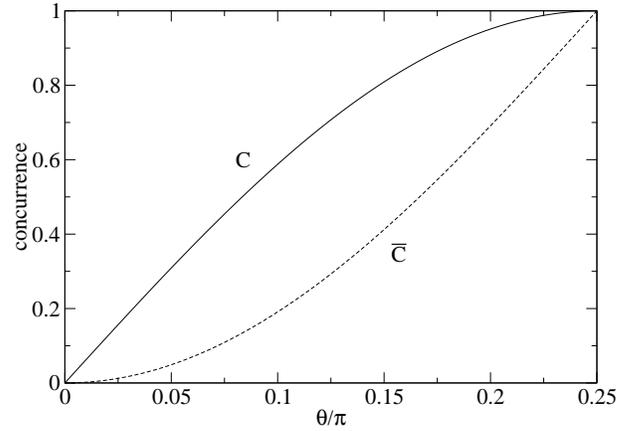}
\caption{The original concurrence $C(t=0)$ (solid line) and final average
concurrence $\bar{C}$(dashed line) as a function of $\theta/\pi$.}
\label{fig:con1}
\end{figure}

{\it Efficiency}: In the above, we have shown that the entanglement of a pair
of the two-qubit system can be enhanced with a certain probability by creating
the coherence in the ancillary degree of freedom and a follow up projection
measurement on them. Now we consider a large number ($N$) of partial entangled
pairs. Obviously we can only obtain a small number of maximally entangled
state, such as singlets, due to its finite probability in the measurement
process (\ref{eq:projectionoperator}). Introducing the final average
concurrence as ${\bar C}=N'/N$ where $N'$ denotes the number of pairs with
maximum entanglement ($C'=1$) after the projection measurement.  Clearly ${\bar
C}$ depends on the value of $\theta$ as was shown in Fig. \ref{fig:con1}, from
which we see that although we can generate state with higher entanglement,
${\bar C}$ still can not exceed $C$ at any value of $\theta$. Moreover, even
though the concurrence is concave function of $\theta$, ${\bar C}$ is not. We
attribute this result to its dependence on the probability of the projection
measurement rather than the initial entanglement itself. We define the
efficiency as $ R\equiv \frac{\bar C}{C} $ where $C=|\sin(2\theta)|$ is the
initial concurrence. For the present case, it is calculated as
$\tan(\theta),(0\leq\theta\leq\pi/4)$.

{\it Discussion and summary:} In this work, we proposed a scheme to control the
entanglement between two separated systems by introducing a local magnetic
pulse and a follow up projection measurement on the ancillary degree of
freedom. Our scheme is quite different from the one proposed by Bennett {\it et
al}, which operates the projection on the whole state of $N$ pairs, while we do
it on individual pairs, so it is easier to be realized by experiment. In our
scheme, the function of the rectangular magnetic pulse is to establish the
coherence between the ancillary degree of freedom, i.e., to transfer partial
population to a target state. Experimentally, this process can be completed by
using adiabatic transfer interferometer\cite{PMarte91,MWeitz94}. Meanwhile, we
would like to point out that the population transfer here is selective, and it
is realized via a magnetic resonance. Therefore, another important feature of
the magnetic pulse is that it act as a filter\cite{DYang04}. The selective
population transfer or the filter can also be realized according to the Pauli
exclusion principle or theory of forbidden band in quantum mechanics (For
example, spin filter in condensed matter physics\cite{PRecher00}).

Moreover, our scheme can be easily generalized to multi-level state as well as
mixed state. Take the latter as an example, if the initial state is $\rho(0)$,
the problem then becomes to find a method to modulate the entry of $\rho(0)$ by
introducing one or more ancillary degree of freedom and a magnetic pulse to
realize the population transfer. That is, during the time evolution, the mixed
state can be written as $\rho(t)=\rho_1(t) + \rho_2(t)$ where $\rho_1(t)$ and
$\rho_2(t)$ are characterized by the ancillary degree of freedom and
$\rho_1(t)$ is assumed to possess higher entanglement then $\rho(0)$. Then
after a projection measurement, the state can be be projected to the desired
state $\rho_1$ with a certain probability.

This work is supported by the Earmarked Grant for Research from the Research
Grants Council (RGC) of the HKSAR, China (Project CUHK 401504). SJGU thanks D.
Yang for helpful and stimulating discussions, We shank S. Y. Zhu for helpful
comments and discussions.


\begin{thebibliography}{99}
\bibitem{Nielsen1}
M. A. Nilesen and I. L. Chuang, {\it Quantum Computation and Quantum
Information} (Cambridge University Press, Cambridge, England, 2000)

\bibitem{AGalindo02}
A. Galindo and M. A. Martin-Delgado, Rev. Mod. Phys. {\bf 74}, 347 (2002).

\bibitem{Bennett}
C. H. Bennett and S. J. Wiesner, Phys. Rev. Lett., {\bf 68}, 557 (1992); C. H.
Bennett, G. Brassard, C. Crepeau, R. Jozsa, A. Peres, and W. Wootters, Phys.
Rev. Lett., {\bf 70}, 1895 (1993).

\bibitem{LMKVandersypen04}
For example, L. M. K. Vandersypen and I. L. Chuang, Rev. Mod. Phys. {\bf 76},
1037 (2004).

%NMR experiments to create entanglement state.
\bibitem{ILChuang98}
N. A. Gershenfeld and I. L. Chuang, Science {\bf 275}, 350 (1997); I. L.
Chuang, {\it et al}, Prog. R. Soc. London, Ser. A {\bf 454}, 447 (1998).

\bibitem{SLloyd98}
S. Lloyd, Phys. Rev. A {\bf 57}, R1473 (1998);

\bibitem{RJNelson00}
R. J. Nelson, D. G. Cory, and S. Lloyd, Phys. Rev. A {\bf 61}, 022106 (2000).


%entanglement creation via adiabatic population transfer
\bibitem{RGUnanyan01}
R. G. Unanyan, N. V. Vitanov, and K. Bergmann, Phys. Rev. Lett. {\bf 87},
137902 (2001); R. G. Unanyan, B. W. Shore, and K. Bergmann, Phys. Rev. A {\bf
63}, 043405 (2001).

\bibitem{BZhou04}
B. Zhou, R. Tao, and S. Q. Shen, Phys. Rev. A {\bf 70}, 022311 (2004).

\bibitem{VSMalinovsky04}
V. S. Malinovsky and I. R. Sola, Phys. Rev. Lett. {\bf 93}, 190502 (2004).


%reference on adiabatic transfer of population
\bibitem{PMarte91}
P. Marte, P. Zoller, and J. L. Hall, Phys. Rev. A {\bf 44}, R4118 (1991).

\bibitem{MWeitz94}
M. Weitz, B. C. Young, and S. Chu, Phys. Rev. Lett. 73, 2563 (1994).


%entanglement distillation
\bibitem{Bennett1}
C. H. Bennett, D.P. DiVincenzo, J. A. Smolin and W. K. Wootters, Phys. Rev. A
{\bf 54}, 3824 (1996).

%entanglement concentration
\bibitem{CHBennett96}
C. H. Bennett, H. J. Bernstein, S. Popescu, and B. Schumacher, Phys. Rev. A
{\bf 53}, 2046 (1996).

%Definition of concurrence.
\bibitem{WKWootters98}
W. K. Wootters, Phys. Rev. Lett. {\bf 80}, 2245 (1998); S. Hill and W. K.
Wootters, Phys. Rev. Lett. {\bf 78}, 5022 (1997).

\bibitem{DYang04}
Dong Yang, Sixia Yu, quant-ph/0410187

%a reference on spin filter
\bibitem{PRecher00}
For example, P. Recher, E. V. Sukhorukov, and D. Loss, Phys. Rev. Lett. {\bf
85}, 1962(2000).

\end{thebibliography}
\end{document}